\documentclass[%
 amsmath,amssymb,
 aps, physrev,showkeys,
]{revtex4-2}

\usepackage{graphicx}
\usepackage{dcolumn}
\usepackage{bm}

\usepackage{xr}
\begin{document}

\preprint{APS/123-QED}

\title{\textbf{Expertise diversity of teams predicts originality and long-term impact in science and technology} 
}%

\author{Weihua Li}
\affiliation{LMIB, NLSDE, BDBC, and School of Artificial Intelligence, Beihang University, Beijing, China}
\affiliation{Department of Advanced Interdisciplinary Research, Pengcheng Laboratory, Shenzhen, China}
\affiliation{Zhongguancun Laboratory, Beijing, China}
\affiliation{Qianyuan Laboratory, Hangzhou, China}

\author{Hongwei Zheng}
\email{Corresponding author: hwzheng@pku.edu.cn}
\affiliation{Beijing Academy of Blockchain and Edge Computing, Beijing, China}

\date{\today}

\begin{abstract}
Despite the growing importance of collaboration networks in producing innovative science and technology, it remains unclear how expertise diversity among team members relates to the originality and impact of the work they produce. Here, drawing on statistical physics, we develop a new computational method to quantify the expertise distance of researchers based on their prior career histories and apply it to 23 million scientific publications and 4 million patents.
We find that across science and technology, teams with expertise diversity tend to produce work with greater originality.
Teams with more diverse expertise exhibit substantially higher long-term impact (10 years), increasingly attracting larger cross-disciplinary influence. This impact premium of expertise diversity among team members becomes especially pronounced when other dimensions of team diversity are missing, as teams within the same institution or country appear to disproportionately reap the benefits of expertise diversity. While gender-diverse teams have relatively higher impact on average, teams with varied levels of gender diversity all seem to benefit from increased expertise diversity. Given the growing knowledge demands on individual researchers, implementation of incentives for innovative research, and the tradeoffs between short-term and long-term impacts, these results may have implications for funding, assembling, and retaining teams with originality and long-lasting impacts.
\end{abstract}

\maketitle


\section{\label{sec:level1}Introduction}

People are increasingly more likely to form collaboration networks in producing innovative work across a wide range of creative domains\cite{wuchty2007increasing,jones2008multi,stokols2008science,fortunato2018science,clauset2017data,wang2021science,dechurch2010cognitive}.
The accumulation of knowledge\cite{jones2009burden} and the specialization of individual researchers\cite{teodoridis2019creativity,schweitzer2021burden} underscore the necessity of forming interdisciplinary teams to tackle complex challenges.
In an era where narrow expertise from a single field may prove insufficient to address pressing societal issues, collaborative efforts of teams spanning traditional disciplinary boundaries become indispensable\cite{guimera2005team,van2005learning,lariviere2010relationship, wang2015interdisciplinarity, gates2019nature}.
Teamwork also allows researchers to establish professional connections, cultivate networks, and engage with a broader audience, both within and beyond their immediate research community\cite{wagner2019international}.
As growing scholarly attention has been paid to increasing diversity, equity, and inclusion in science, the scientific community demands more efforts to improve representational diversity in the composition of research teams\cite{casad2021gender,ncs2024editorial}.
A better understanding of how to assemble teams with diverse expertise is therefore essential for coordinating collective actions, fostering interdisciplinary thinking, and integrating existing expertise to tackle new challenges across scientific and technological domains\cite{van2011factors,lungeanu2014understanding,mayrose2015interplay,hicks2015bibliometrics,hall2018science,smith2021great,zeng2022impactful}.

However, it remains unclear how prior expertise diversity among team members is related to the originality and the impact of the work the team produces.
Some indicators have focused on quantifying the diversity of the prior expertise among individuals, by calculating for example the Pearson correlation of the distribution across technological classes distributions\cite{nooteboom2007optimal},
a research overlap score using medical subject headings (MeSH) terms\cite{mayrose2015interplay},
a frequency-inverse document frequency (TF-IDF) method for research content similarity\cite{araki2017interdisciplinary},
latent semantic analysis for the topic similarity between researchers\cite{smith2021great}, a cosine distance estimate for prior expertise disparity\cite{hill2021adaptability}, and a distance metric based on the Jaccard dissimilarity of researchers' references\cite{zeng2021fresh}.
Although these methods can approximate expertise diversity among team members, an effective measure should also explicitly account for the relatedness of research disciplines\cite{stirling2007general,lariv2015long}.
For instance, ecology has more interactions with environmental science and evolutionary biology than condensed matter physics.
Similarly, artificial intelligence is more deeply influenced by mathematics and computer science compared to organic chemistry.
Therefore, ecologists should on average have higher expertise similarity to environmental scientists and evolutionary biologists than condensed matter physicists, and the knowledge possessed by artificial intelligence researchers may be more related to the knowledge of mathematicians and computer scientists than that of organic chemists.
	
In part to tackle these challenges, researchers have developed several methods to infer the diversity of knowledge scope from the product the team has developed, by using a paper's references or citations, using measurements such as the distinction of fields, entropy, Gini coefficient, and the Rao-Stirling (RS) index, to quantify interdisciplinarity of the work and its association with impact\cite{stirling2007general,yegros2015does,waltman2015field,wang2017bias,leydesdorff2019interdisciplinarity,yu2021become}.
While these measures allow researchers to quantify the interdisciplinarity of the work and its association with originality and impact, as proxies for knowledge diversity, they depend upon the paper that a team has produced, by analyzing its references or citations, which are only available after the fact, i.e., after the paper has been published.
Moreover, these metrics are designed to quantify the diversity of knowledge scope implemented in a specific piece of work, rather than a comparative measure to estimate the disparity of expertise among individual scholars.
It remains unclear how the composition of team expertise is related to the originality and impact of research a team is about to produce.
Understanding the association between the expertise diversity among team members and the outcome the team produces is crucial for funding and investment decisions, and highlights the importance of developing measures to quantify and combine diverse prior knowledge to inform fruitful collaboration strategies.

To address these challenges, we propose a new metric to identify and quantify the diversity of prior expertise among team members that extends beyond single disciplines. Our expertise distance metric explicitly accounts for the relatedness of scientific fields and draws on the disciplinary distributions of prior career histories among collaborators. The expertise diversity of a team is then obtained as the average distance of all possible pairwise coauthorships, which correlates with a broad range of indicators of scholarly diversity in terms of the combination of past knowledge, and other dimensions of diversity among team members regarding their affiliations, nationality, and gender.
	
A large body of work has focused on understanding the interplay between team composition and outcomes, probing dimensions of team diversity around affiliations\cite{jones2008multi}, ethnicity\cite{alshebli2018preeminence}, gender\cite{nielsen2017opinion}, expertise\cite{van2005learning}, technical background\cite{jackson1996consequences}, problem-solving ability\cite{hong2004groups}, intelligence\cite{woolley2010evidence}, and more. In the technology sector, inventors from distinct social groups tend to generate patents with greater collaborative creativity\cite{fleming2007collaborative}. In the online knowledge-sharing community, polarized teams composed of a balanced proportion of ideologically diverse Wikipedia editors produce articles of higher quality than homogeneous teams\cite{shi2019wisdom}. 
Some scholars have found that multidisciplinarity exhibits an inverted U-shape relationship with scientific impact, although the vast majority of papers indicate a positive correlation with multisciplinarity\cite{song2023influence}. 
Furthermore, the value of multidisciplinary research might require a longer period to be recognized by the scientific community, leading to delayed impact\cite{Zhang2022DelayedIO}.
Overall, studies from diverse domains have demonstrated that the impact of team outcomes improves with team diversity. Thus, creative output in science and technology may be heavily rooted in the composition of team members' expertise and background, which is largely determined during the team assembly process.

Originality is often regarded as a core goal in science and technology. Recent work has suggested that creative ideas often emerge from new and unconventional combinations of knowledge from diverse disciplines, research methods, or frameworks\cite{uzzi2013atypical,wagner2019international}.
Innovation can be spurred when proven methodologies in one domain are introduced to solve problems in a fresh area\cite{Guimer2005TeamAM}.
Creativity is more likely to emerge from diverse teams as researchers integrate concepts and methods drawn from diverse disciplines, forging connections between seemingly disparate concepts or bodies of knowledge\cite{wuchty2007increasing,Schilling2011RecombinantSA}.
Although studies have suggested that high multidisciplinarity is associated with high impact, our findings suggest that scientific teamwork with multidisciplinary approaches has little correlation with originality, quantified by the disruption score, and has negative correlations with originality in technology.
Research by expertise-diverse teams, in contrast, is positively correlated with high originality in science and technology.
Therefore, despite its close relation with multidisciplinarity, the expertise diversity of teams not only presents a new quantitative measure to understand science but also provides new perspectives for the originality of team outcomes.
	
Moreover, while scholars have argued that multidisciplinarity is associated with the impact of research\cite{Leahey2017ProminentBL,Chen2021ExploringTI}, we find that research teams with high expertise diversity exhibit no significant impact advantage in the short- (2 years) or mid-term (5 years).
This pattern persists for teams spanning both scientific and technological domains. We find that, instead, teams with high expertise diversity enjoy a substantive impact premium of their work in the long-term (10 years), increasingly attracting cross-disciplinary influence in the longer run.
The long-term effect of expertise diversity becomes more prominent as team size and citation time window grow.
In particular, when other dimensions of diversity are missing, teams formed in the same institution or country disproportionately harness the benefit of expertise diversity. These results may have implications for fostering and retaining innovative teams with more diverse knowledge composition among team members.

\section{Expertise distance metric}
We use the field distribution of the publication history of individual authors to quantify the expertise distance among coauthors.
Suppose for field $f$, the cumulative reference vector $\boldsymbol{v_f} = (v_f^1, ..., v_f^n)$ records the numbers of references of papers from field $f$ made to all fields, where $n$ is the number of fields\cite{stirling2007general, gates2019nature}. We then define the unit vector of field $f$ as $\boldsymbol{e_f} = \boldsymbol{v_f} / ||\boldsymbol{v_f}||$, where $||\boldsymbol{v_f}|| = \sqrt{\sum_m (v_f^m)^2}$.
	
For an author $i$, let the publication vector $\boldsymbol{p_i}$ record the number of publications author $i$ had in each field up to year $t$.
The goal is to introduce a definition of expertise distance between a pair of authors that accounts for the relatedness between fields.
An explicit form of $\boldsymbol{p_i}$ is $\boldsymbol{p_i} = a_i^1 \boldsymbol{e_1} + ... + a_i^f \boldsymbol{e_f} + ... + a_i^n \boldsymbol{e_n}$, where $a_i^f$ is the number of papers author $i$ published in field $f$.
We want to first normalize $\boldsymbol{p_i}$ and obtain expertise vector $\boldsymbol{q_i}$ of unit length $\boldsymbol{q_i}=\boldsymbol{p_i}/||\boldsymbol{p_i}|| = q_i^1 \boldsymbol{e_1} + ... + q_i^f \boldsymbol{e_f} + ... + q_i^n \boldsymbol{e_n}$ and $\overline{\boldsymbol{q_i}} = (q_i^1, ..., q_i^f, ..., q_i^n)$.
Note that different from the definition of field vector length, here we explicitly account for the heterogeneous relatedness between fields and define
\begin{equation}
	||\boldsymbol{p_i}||^2 = (a_i^1 \boldsymbol{e_1} +  ... + a_i^f \boldsymbol{e_f}+ ... + a_i^n \boldsymbol{e_n})^2 = \sum_{j,k} a_i^j a_i^k (\boldsymbol{e_j} \cdot \boldsymbol{e_k}).
	\label{normalize1}
\end{equation}
	
Similarly, the cosine distance between expertise vectors of authors $i$ and $j$ can be calculated as $\boldsymbol{q_i} \cdot \boldsymbol{q_j} = \overline{\boldsymbol{q_i}} \boldsymbol{M} \overline{\boldsymbol{q_j}}^T$, and the distance metric can be obtained as
\begin{equation}
	d_{ij} = \sqrt{|| \boldsymbol{q_i} - \boldsymbol{q_j}|| ^ 2} = \sqrt{2 - 2 \boldsymbol{q_i} \cdot \boldsymbol{q_j}} = \sqrt{2 - 2 \overline{\boldsymbol{q_i}} \boldsymbol{M} \overline{\boldsymbol{q_j}}^T},
	\label{distance}
\end{equation}
where $\boldsymbol{M}=(\boldsymbol{e_i}\boldsymbol{e_j})_{i,j}$ is the matrix indicating the closeness between fields using cosine distance.
Analogously, if we let $\overline{\boldsymbol{p_i}} = (p_i^1, ..., p_i^f, ..., p_i^n)$, the length of publication vector of author $i$ can be obtained as
\begin{equation}
	||\boldsymbol{p_i}|| = \sqrt{\overline{\boldsymbol{p_i}} \boldsymbol{M} \overline{\boldsymbol{p_i}}^T}.
	\label{normalize}
\end{equation}
	
To estimate the prior expertise distance of a pair of authors $(i,j)$ that coauthored a paper $s$ at year $t$, we first build their respective publication vectors using all papers up to year $t$.
We use the level 1 field classification in the MAG dataset, where each article is assigned to at least one scientific field.
For the crude publication vector $p_i(t)$ of author $i$, $a_i^f$ denotes the number of papers $i$ published in the field $f$ up to $t$.
To account for authors with a reasonably long publishing career, we include only productive authors who have published at least 5 papers up to $t$.
The mean expertise distance of the paper $\langle d_s \rangle$ is the average distance of all possible coauthor pairs $(i,j)$ among selected productive authors.
	
Papers with at least two productive authors are eligible to obtain a measure of team expertise distance, and those written by solitary authors are not considered in this study.
To allow for a meaningfully long prior publication record of individual researchers, we consider only authors who have published at least 5 papers by the time of collaboration when estimating the expertise distance of the team.
Thus, a proportion of less productive or early-career researchers are dropped and we obtain a subset of team-authored papers eligible for a distance metric.

\section{Data}
We use the publication and citation data in 1950-2019 from the Microsoft Academic Graph (MAG) dataset\cite{wang2020microsoft}.
We select papers published in journals for science, technology, and social sciences, namely biology, business, chemistry, computer science, economics, engineering, environmental science, geography, geology, materials science, mathematics, medicine, physics, political science, sociology, and papers published in conferences for computer science.
We also extract patent data from the MAG archive.

There are some known limitations and deficiencies of MAG, particularly regarding citation data\cite{liang2021finding}. To address this issue, we implement a rigorous data filtering process before the analysis (see also Supplementary Information). In particular, we exclusively include papers sourced from journals or conference proceedings within MAG, and ensure that all papers contain author affiliation information. 
After data filtering procedures, we retain a total number of 22.8 million papers, and 354 million citations from papers to papers. We also extract 4.4 million patents, which make 29.9 million citations among themselves, and 5.6 million citations to the selected subset of research articles.

\section{Results}

\begin{figure*}[htbp]
\centering
\includegraphics[width=0.9\linewidth]{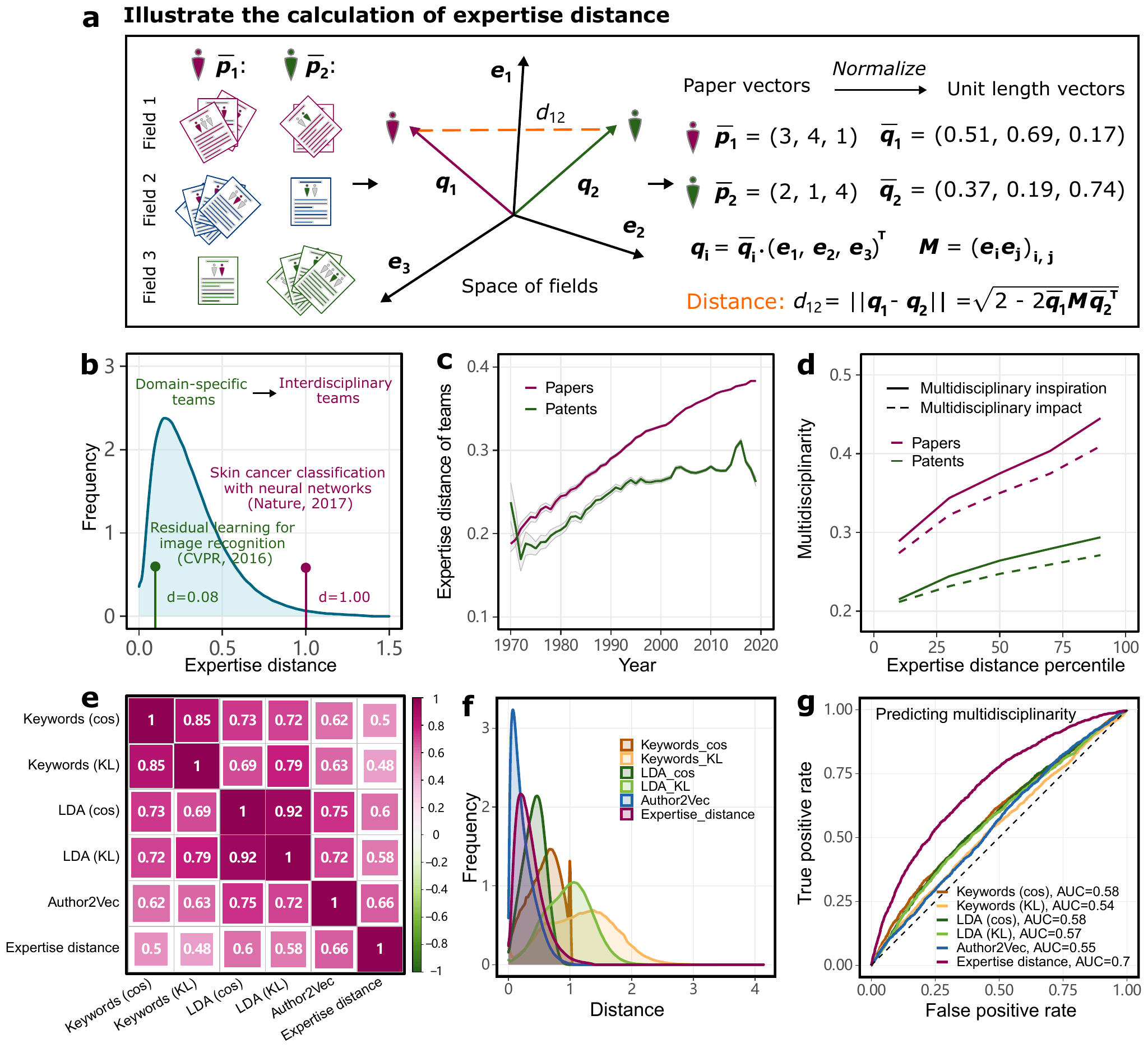}
\caption{\footnotesize \textbf{Illustrating the estimation of expertise distance among team members and its correlation with multidisciplinarity.}
(\textbf{a}) We first indicate the prior career histories of two authors $1$ and $2$, distributed across three research fields $1$, $2$, and $3$.
Then we display the expertise vectors $\boldsymbol{q_1}$ and $\boldsymbol{q_2}$ and their prior expertise distance $d_{12}$ in the geometric space of research fields, in which the field vectors $\boldsymbol{e_j}$ are usually non-orthogonal to each other due to the heterogeneity of interactions among fields, where $j \in (1,2,3)$.
To estimate the expertise distance, we obtain the expertise vectors $\overline{\boldsymbol{q_i}}$ of unit length $1$ by normalizing the publication vectors $\overline{\boldsymbol{p_i}}$ of authors, where $i \in (1,2)$.
The expertise vector $\boldsymbol{q_i}$ shown in the space of fields can be related to the expertise vector $\overline{\boldsymbol{q_i}}$ as $\boldsymbol{q_i} = \overline{\boldsymbol{q_i}} (\boldsymbol{e_1}, \boldsymbol{e_2}, \boldsymbol{e_3})^T$.
We then compute the coauthor expertise distance $d_{12}$ by taking into account the interaction matrix $\boldsymbol{M}$ that explicitly embeds in the relatedness of fields.
(\textbf{b}) The distribution of the average expertise distance among team members for over 11 million research papers published in 1970-2019. Then we mark the skin cancer classification paper using neural networks ($d = 1.00$, top $2\%$) that is more interdisciplinary in terms of both the composition of team expertise and the produced work, and the image recognition paper ($d=0.08$, bottom $5\%$) that is conducted by a team with members focusing primarily on a specific research area. Both papers are highly cited.
(\textbf{c}) Contemporary trends of teams' expertise diversity for papers and patents, from 1970 to 2019. Shaded areas represent 95\% confidence intervals.
(\textbf{d}) To validate the efficacy of the distance metric, we show the interplay between teams' expertise distance percentile and the multidisciplinary inspiration (solid lines) and impact (dashed lines) of papers and patents.
(\textbf{e}) We operationalize several alternative methods to assess expertise and show the correlation matrix between them and the expertise distance metric.
(\textbf{f}) Probability distribution of all expertise metrics.
(\textbf{g}) ROC curves of different expertise metrics predicting multidisciplinarity of teams' work.
}
\label{illustrate}
\end{figure*}

We propose a new metric to measure the expertise distance between two researchers based on their prior career histories, which we illustrate in a concise example (Fig.~\ref{illustrate}).
The publication vector $\overline{\boldsymbol{p_i}}$ records the crude number of papers by the author $i$ distributed across all research fields at the year of collaboration.
We define $\overline{\boldsymbol{p_1}} = (3, 4, 1)$ and $\overline{\boldsymbol{p_2}} = (2, 1, 4)$ in the illustrative model for the selected authors $1$ and $2$, respectively (Fig.~\ref{illustrate}{a}).
The prior expertise vectors $\boldsymbol{q_1}$ and $\boldsymbol{q_2}$ with unit length are shown in the geometric space of research fields $(\boldsymbol{e_1}, \boldsymbol{e_2}, \boldsymbol{e_3})$, obtained by normalizing the raw publication vectors $\overline{\boldsymbol{p_1}}$ and $\overline{\boldsymbol{p_2}}$.
The field vectors in this space are usually not orthogonal to each other, due to heterogeneity across fields.
Both the vector normalization and the distance estimation take into account the field interaction matrix $\boldsymbol{M}$, where each element $m_{ij}$ is calculated as the dot product of the corresponding field vectors $\boldsymbol{e_i}$ and $\boldsymbol{e_j}$ of field $i$ and $j$ (see Methods).
The final analytical form of author expertise distance is given in Eq.~(\ref{distance}).
As such, the derived distance metric varies between $0$ and $\sqrt{2}$, and a higher distance value corresponds to greater prior career diversity between the two authors.
The expertise diversity of a team is therefore defined as the average expertise distance among all possible pairs of coauthors.

Empirical evidence suggests that teams of either high or low expertise diversity may possess the potential to develop original and high-impact research.
However, the level of interdisciplinarity in the approaches implemented by high-originality or high-impact teams varies substantively, which can be predicted by the team members' expertise diversity estimated using the proposed metric.
For instance, a team that applied a deep neural network method to classify skin cancer involves researchers from departments of electrical engineering, dermatology, microbiology \& immunology, and computer science, spanning a wide range of research areas (Fig.~\ref{illustrate}{b})\cite{esteva2017dermatologist}.
Correspondingly, they proposed a highly interdisciplinary approach by creatively adapting artificial intelligence technologies to life science problems in the paper.
In another example, a group of researchers from the same institution who focus primarily on computer vision produced a well-cited paper that trains deep residual learning networks for image recognition\cite{he2016deep}.
Their paper presents a new framework within a particular research topic of computer vision.
The distance metric accurately captures the two teams' propensity to conduct interdisciplinary research based on the prior career histories of team members.

In the following analyses, we aggregate groups of research papers according to the sizes of teams, defined by the number of authors of a paper.
Previous studies show that a team's preferences for research agenda, originality, and expected impact are strongly associated with its size.
Large teams tend to build their work on more recent developments and have a higher impact, whereas small teams are more likely to disrupt science and technology\cite{wu2019large,wuchty2007increasing}.
Aggregating teams by their sizes mitigates this confounding effect and controls for the structural variations of the coauthorship network of teams with varying sizes.
As such, we categorize teams of the same size into five uniform percentile bins based on the team members' average expertise distance in our analyses.

\begin{figure*}[htbp]
\centering
\includegraphics[width=0.6\linewidth]{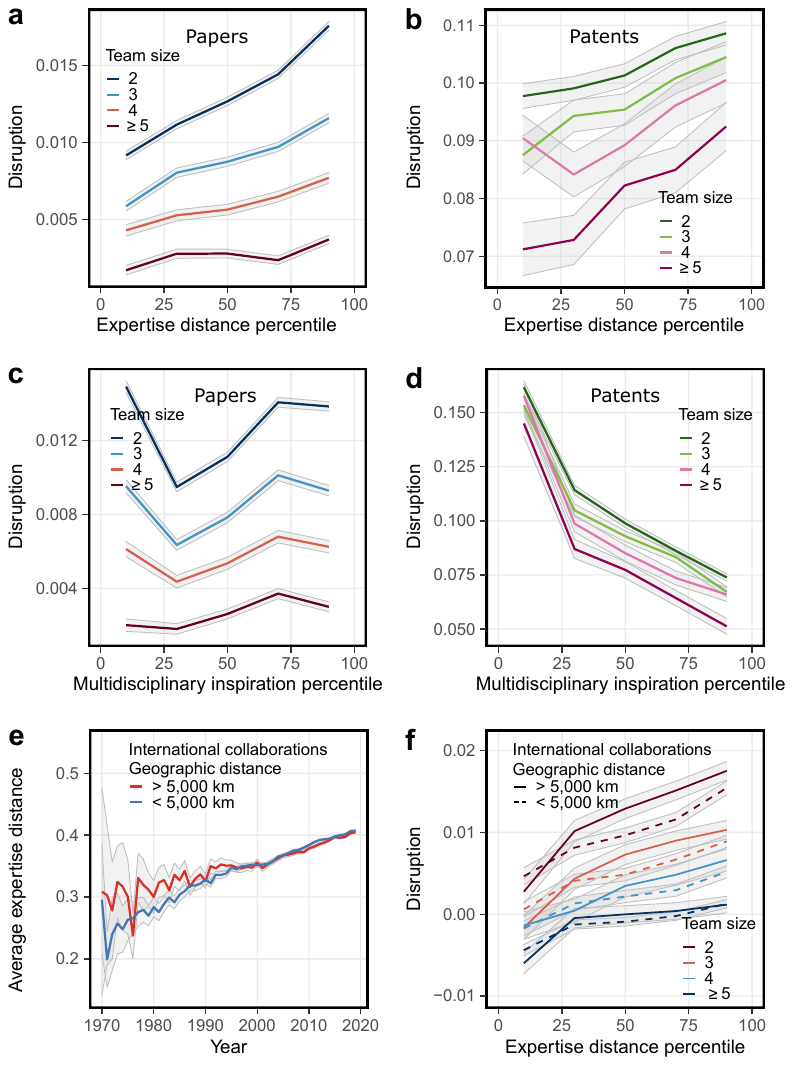}
\caption{\footnotesize \textbf{Originality and its correlation with expertise diversity of teams and multidisciplinarity of the teams' work.}
We present the interplay between expertise diversity of teams and disruption of the teams' work and find that high expertise diversity is correlated with a high disruption score for both papers (\textbf{a}) and patents (\textbf{b}).
In contrast, we show the interplay between multidisciplinary inspiration and disruption of the teams' work. We find that multidisciplinary inspiration has no consistent correlation with disruption for papers (\textbf{c}), and appears to be negatively correlated with disruption for patents (\textbf{d}).
(\textbf{e}) Contemporary trends of expertise diversity and geographic distance of international teams from 1970 to 2019. 
(\textbf{f}) Correlation between the average geographic distance and disruption of research by international teams.
Shaded areas represent 95\% confidence intervals.
}
\label{dist_disruption}
\end{figure*}

We extract and refine 23 million original research articles and over 4 million patents from the Microsoft Academic Graph (MAG) dataset in the period of 1950 to 2019,
and we use the level 1 field annotations of articles provided by MAG to quantify past research experience and calculate expertise diversity.
At the level of coauthor pairs, expertise distance follows a log-normal distribution for both papers and patents, and declines as the duration of collaboration extends.
The expertise diversity of scientific research teams has been continually increasing over time since the 1970s, suggesting that teams have become more diverse in the distribution of knowledge composition among the members (Fig.~\ref{illustrate}{c}).
The contemporary trend of teams' expertise diversity also varies across disciplines.
Research teams in natural and social sciences have become substantially more diverse over time, whereas the historical level of expertise diversity for engineering \& mathematics teams has been fairly stable in the past decades.
	
To validate whether our distance metric accurately captures the diversity of knowledge composition among team members, we measure the diversity of field distributions in the references or citations of papers and patents using the Rao-Stirling index and see how it relates to team expertise diversity\cite{stirling2007general}.
An RS index of zero means the work's references or citations concentrate mostly on one particular field, while an RS index close to one means the work's references or citations distribute broadly across many fields.
The multidisciplinary inspiration of papers quantified by the RS index using references increases as the teams' expertise diversity grows, suggesting that high-distance teams tend to reference a wider range of disciplines than low-distance teams (Fig.~\ref{illustrate}{d})\cite{gates2019nature}.
Similarly, the multidisciplinary impact of papers measured by the RS index using citations is also positively correlated with the teams' expertise diversity, suggesting that works of high-distance teams tend to have broader cross-disciplinary influence.
We also validate this by replicating the analysis using the DIV* multidisciplinary indicator \cite{leydesdorff2019diversity,rousseau2019leydesdorff}.

Applying our metric to patenting, we find that expertise diversity of teams has grown modestly over the past decades.
A team's expertise diversity is positively correlated with the multidisciplinary inspiration and multidisciplinary impact of the patent it produced.
These results exhibit the close coupling between expertise diversity and multidisciplinarity, which validates the efficacy of the expertise distance metric for both science and technology.

To further validate the robustness of the expertise distance metric, we operationalize several alternative methods to assess expertise distance.
These methods include utilizing keywords, latent Dirichlet allocation (LDA)\cite{blei2003latent,griffiths2004finding}, and Author2Vec\cite{ganesh2016author2vec,maharjan2019jointly,wu2020author2vec}, and we use cosine distance and Kullback-Leibler (KL) divergence to assess the similarity between the two vectors. 
The metrics demonstrate strong correlations with each other (Fig.~\ref{illustrate}{e}).
Despite varying value ranges among these metrics, they all exhibit a log-normal distribution (Fig.~\ref{illustrate}{f}).

To evaluate the performance of expertise metrics, we employ them in predicting the multidisciplinary inspiration and impact of research papers and visualize the results using ROC curves (Fig.~\ref{illustrate}{g}). 
We find that keywords-based, LDA-based, and Author2Vec metrics exhibit weaker correlations with multidisciplinary inspiration, as evidenced by AUC scores ranging from 0.54 to 0.58. In contrast, expertise distance proves to be more predictive of multidisciplinary inspiration, exhibiting a significantly enhanced AUC score of 0.7.
This result suggests that the expertise distance metric better captures the multidisciplinarity of team outcomes.

Disruption is a metric used to gauge the level of originality of research, ranging from $-1$ to $1$\cite{funk2017dynamic,wu2019large,zeng2021fresh,park2023papers,leibel2024we}.
A paper is regarded as more disruptive and original if a high proportion of citations it receives do not cite its references, indicating that these follow-up studies appear to be primarily inspired by this particular research rather than the body of work it builds on.
We find that expertise diverse teams are more likely to perform disruptive research in science (Fig.~\ref{dist_disruption}{a}), especially for small teams.
High expertise distance teams are also associated with high disruption in technology (Fig.~\ref{dist_disruption}{b}), and the effect is particularly prominent for large teams with more than 5 authors.
The positive correlation between expertise diversity and disruption is consistent across disciplines and decades, and using other variants of the disruption indicator\cite{bornmann2020disruptive,leydesdorff2021disruption}. 
These results suggest that expertise diversity correlates positively with the disruption of teams for science and technology.
			
In contrast, the multidisciplinary inspiration of a paper reflects the disciplinary diversity of its references and approaches utilized in the study.
This indicator can only be calculated when the work is finished, lagging the availability of expertise diversity among team members which is available when the team is assembled.
However, we observe no consistent pattern between multidisciplinary inspiration and its association with disruption in science (Fig.~\ref{dist_disruption}{c}).
Conversely, the multidisciplinary inspiration of patents is negatively associated with disruption, suggesting that highly disruptive patents are rather likely to focus on narrow research areas (Fig.~\ref{dist_disruption}{d}).
These patterns between expertise diversity and disruption are consistent across disciplines and decades, which demonstrates the unique power of team expertise diversity for predicting originality that is not observed in multidisciplinarity, compared to other known measures of research diversity.

Recent studies have examined the disruption of papers in international collaborations \cite{wagner2019international,wagner2019international2,lin2023remote}, and we contribute to this effort by exploring the role expertise diversity plays in shaping the originality of international team outcomes. We divide international teams into two groups based on whether the average geographic distance between coauthor affiliations exceeds 5,000 km.
We find that the geographic distance has little effect on the expertise diversity of international teams over time(Fig.~\ref{dist_disruption}{e}).
In terms of team output, increasing expertise diversity enhances the disruption of scientific research for both groups of international teams, regardless of their geographical distance(Fig.~\ref{dist_disruption}{f}).
This finding may imply that although research by international teams may exhibit lower levels of creativity compared to domestic teams\cite{lin2023remote}, increasing expertise diversity among team members could be crucial for generating more innovative work.

\begin{figure*}[htbp]
\centering
\includegraphics[width=0.8\linewidth]{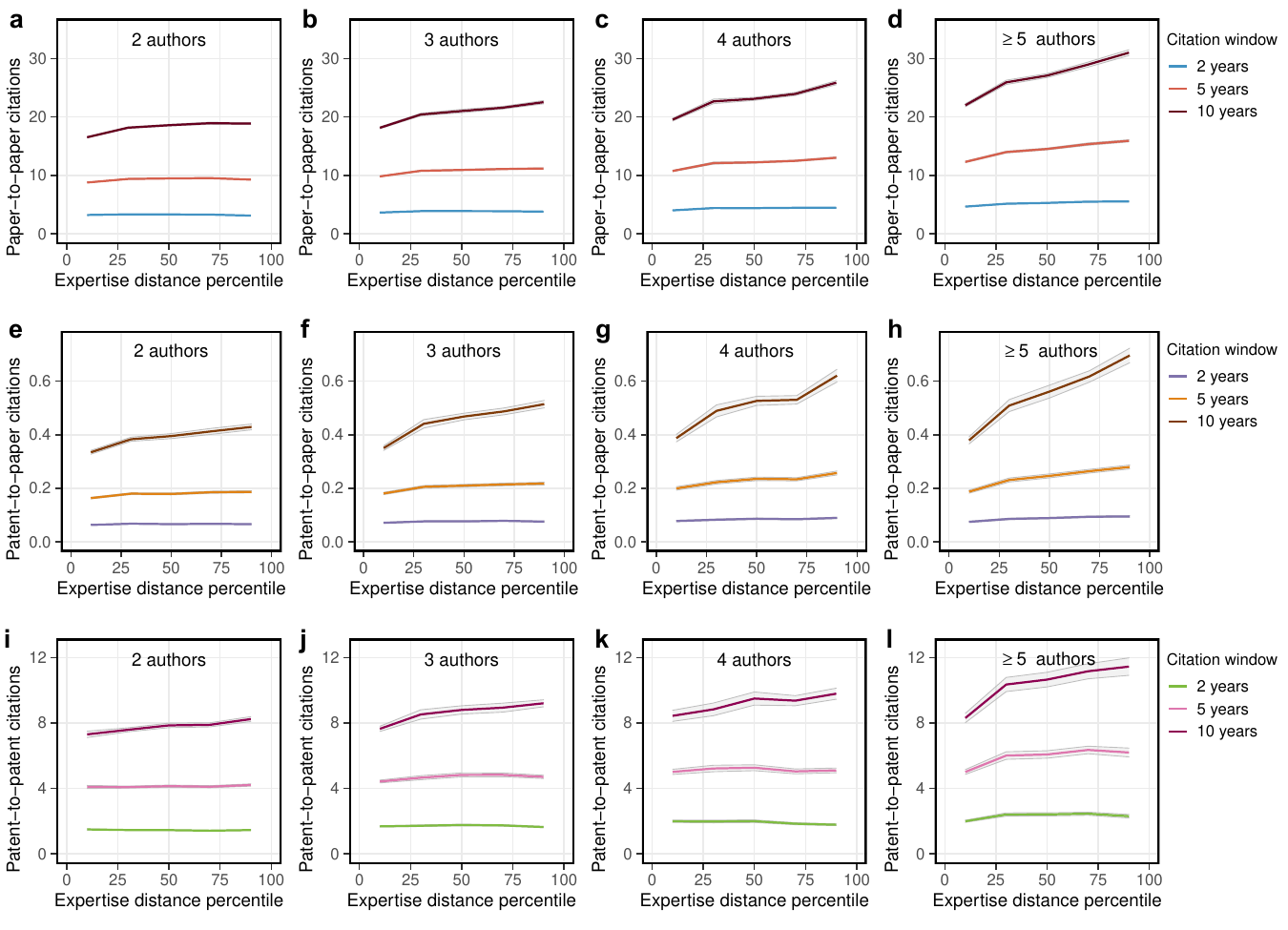}
\caption{\footnotesize \textbf{Distance metric and impact in science and technology.}
We show the interplay between research impact and the teams' expertise diversity divided into five uniform distance percentile bins. We consider three different lengths of citation time windows of two years, five years, and ten years after publication.
We identify three existing citation patterns among science and technology: (\textbf{a-d}), paper-to-paper citations; (\textbf{e-h}), patent-to-paper citations; and (\textbf{i-l}), patent-to-patent citations.
For each pattern, results are presented in separate panels categorized by the teams' size. Shaded areas represent $95\%$ confidence intervals.
}
\label{p_dist_cits}
\end{figure*}

We then proceed to explore other perspectives of team performance related to the expertise diversity among team members.
High-distance teams, composed of researchers with a broad spectrum of expertise, exhibit greater potential to apply more novel combinations of past knowledge, draw from more diverse fields, and produce more innovative results.
Recent studies suggest that interdisciplinarity in scientific research may withstand underestimated short-term impact, but is usually rewarded with high impact in the longer run\cite{wang2017bias}.
We identify and examine three widely existing citation patterns across science and technology to quantify different types of impact.
The first is the most commonly used impact measure of research papers, which we refer to as the paper-to-paper citations (Fig.~\ref{p_dist_cits}{a-d}).
Moreover, studies have shown that successful scientific publications are more likely to inspire future technological innovations, which we use the number of patent-to-paper citations as an indicator to reflect the influence of scientific research on patenting (Fig.~\ref{p_dist_cits}{e-h})\cite{ahmadpoor2017dual}.
Lastly, we measure the impact of patents using the number of patent-to-patent citations (Fig.~\ref{p_dist_cits}{i-l}).
	
For both papers and patents, we find that a team's expertise diversity has no significant correlation with the short-term (2 years) impact of their work, quantified by the number of citations received two years after publication, regardless of its size and type of impact (Fig.~\ref{p_dist_cits}).
For teams with relatively large sizes of four or more authors, high-distance teams have a moderately larger mid-term (5 years) impact than low-distance teams in all three citation patterns, and the effect is less significant for small-sized teams of two or three authors.
Teams of all sizes exhibit substantively greater long-term (10 years) impact in all citation patterns as the teams' expertise diversity increases, and this effect becomes more prominent for large teams.
In particular, papers published by teams with five or more authors in the highest distance percentile bin garner on average $31.0$ paper-to-paper citations in 10 years, $40.9\%$ more than the citations of papers produced by teams from the lowest distance percentile bin (Fig.~\ref{p_dist_cits}{d}).
The effects persist when we remove all self-citations.
The long-term impact premium for expertise-diverse teams still exists when we use a dichotomous variable of highly-cited papers and patents to indicate whether they received the upper 5th percentile of citations for a given year and field.
We further validate the results in two extended citation time windows of 20 and 30 years for papers and patents published in the 1980s and find that the impact premium of high-distance teams becomes even more pronounced.
	
We further test the robustness of the above findings according to the publication period and research discipline of papers and patents.
For all three types of impact, greater team expertise diversity is consistently associated with higher long-term impact across decades, when we independently examine papers and patents published in the 1980s, 1990s, and 2000s. 
Regarding research disciplines, however, more fluctuations exist across these patterns of impact.
For all three types of impact in natural sciences, larger expertise distance of teams is consistently associated with higher long-term impact, while such correlation is not significant and sometimes negative for engineering \& mathematics. 
For large social science teams, greater expertise diversity is related to higher long-term paper-to-paper and patent-to-paper citations, but no identifiable pattern in patent-to-patent citations as relatively fewer patents are issued in social sciences.
	
\begin{figure*}[htbp]
\centering
\includegraphics[width=0.8\linewidth]{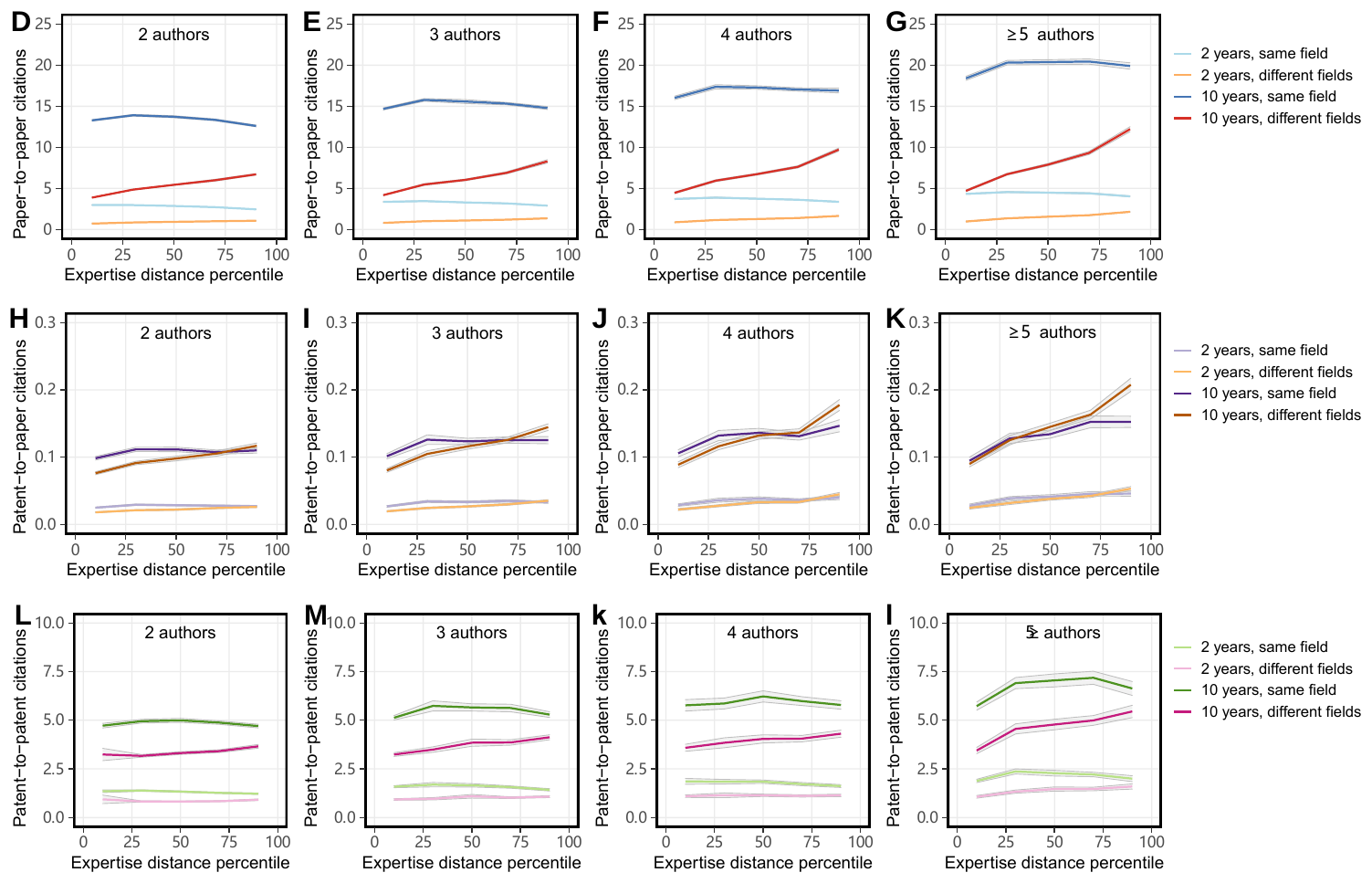}
\caption{\footnotesize \textbf{Teams' expertise diversity and origins of citations in science and technology.}
We decompose the citation origins of fields for scientific and technological research according to the teams' expertise diversity.
We compare the number of citations coming from the same field or from different fields, accrued within two different time windows of 2 years and 10 years after publication, respectively.
Consistent with previous results, we use three citation patterns among science and technology: (\textbf{a-d}), paper-to-paper citations; (\textbf{e-h}), patent-to-paper citations; and (\textbf{i-l}), patent-to-patent citations.
For each pattern, results are presented in separate panels categorized by the team's size. Shaded areas represent $95\%$ confidence intervals.
}
\label{cit_diffFld}
\end{figure*}

To investigate the mechanisms underlying why high-distance teams outperform low-distance teams in the long term but not in the short run, we probe into the disciplinary origins of citations.
Previous studies have suggested an audience effect, whereby coauthors with more diverse networks are more likely to attract scholarly attention to their research\cite{wagner2019international}.
Because knowledge-diverse teams are more likely to draw attention and influence research beyond the constraints of disciplines, we expect that high-distance teams may have a higher cross-disciplinary impact.
We distinguish two citation patterns according to the origins, the number of citations a paper received from papers in the same field, and the number of citations from papers in other fields.
We find that teams of different levels of expertise diversity have similar scientific impact within the field (Fig.~\ref{cit_diffFld}{a-d}), an effect that is persistent for both the short- (2 years) and long-term (10 years).
While high-distance teams do not exhibit a strong impact advantage on different fields in the short term, they attract significantly more citations from other fields in the long term, which becomes more prominent as team size grows.
This suggests that while disciplinary impact consistently exhibits little variance for scientific teams, works of diverse teams attract more cross-disciplinary citations, resembling rippling dynamics that accelerate gradually over time.
	
The long-term boosting effect of cross-disciplinary influence associated with teams' expertise diversity appears to extend beyond science and persists in the technology sector.
Papers produced by high-distance teams have higher cross-disciplinary impact on patented inventions in the long term, but less so in the short run (Fig.~\ref{cit_diffFld}{e-h}).
For patents, the cross-disciplinary impact advantage for high-distance teams is relatively weak compared to that of papers, but still with similar boosting effects in the long term (Fig.~\ref{cit_diffFld}{i-l}).
These results show that the long-term impact premium of high-distance teams in science and technology is largely driven by the cross-disciplinary attention progressively garnered over time.
	
\begin{figure*}[htbp]
\centering
\includegraphics[width=0.8\linewidth]{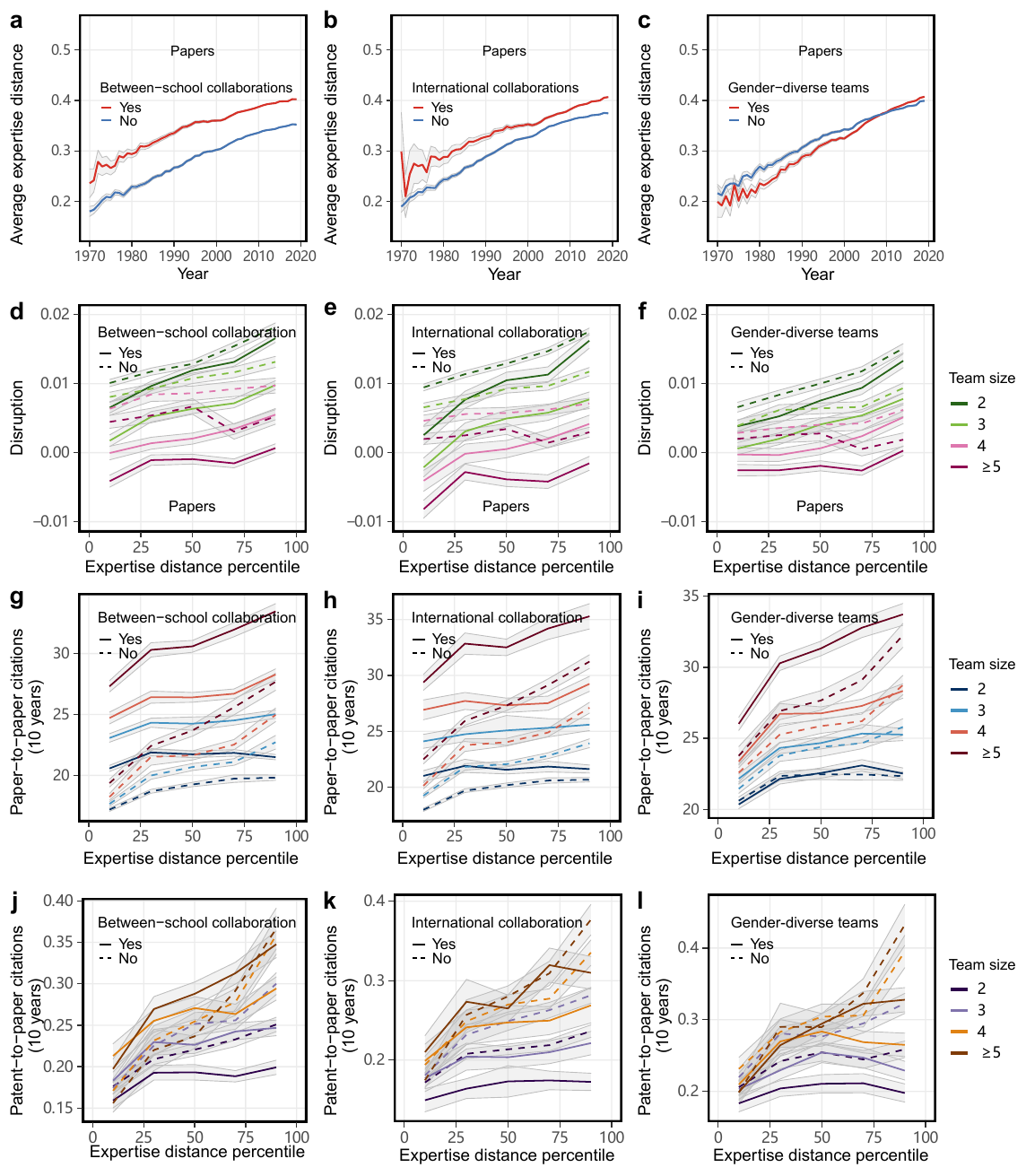}
\caption{\footnotesize \textbf{The synthesized effects of teams' expertise diversity and other dimensions of team diversity on long-term impact.}
We present the temporal trends of team expertise diversity and its correlation with disruption and two types of impact in the long run, conditional upon other dimensions of diversity among team members including affiliations, nationality, and gender.
(\textbf{a-c}) Temporal trends of team expertise diversity based on the diversity of team members' affiliations, nationality, and gender.
The effect of team diversity and expertise diversity on (\textbf{d-f}) disruption, (\textbf{g-i}) paper-to-paper citations, and (\textbf{j-l}) patent-to-paper citations.
In particular, we consider three types of diversity among team members.
Left column, whether it is a single-institutional or between-school team.
Middle column, whether the team involves international collaborators.
Right column, whether the team is gender-diverse or gender-homogeneous.
Teams with diverse background collaborators, i.e., having between-school collaborations, having international collaborations, or having female researchers on board, are indicated by solid lines. Teams with simple background collaborators are shown in dotted lines. Shaded areas represent $95\%$ confidence intervals.
}
\label{cit_diversity}
\end{figure*}
	
We further disaggregate team expertise diversity based on the background diversity of team members in terms of affiliations, nationality, and gender\cite{page2008difference}.
Given that the vast majority (over $99\%$) of patents are registered by single-institutional teams, here we focus on research papers only in our analyses.
Scientists tend to collaborate more with people in their own groups\cite{bozeman2004scientists}, and are less inclined to form cross-disciplinary teams with people from more remote positions\cite{vacca2015designing}.
Therefore, the expertise distance of teams may be associated with the geographical proximity of researchers' affiliations and locations\cite{jones2008multi}.
Teams involving between-school or international collaborations are likely to encompass a more diverse composition of knowledge, expertise, and skills than teams assembled within the same institution or country.
We find that the team's expertise diversity is positively correlated with the propensity of having collaborators from external or international institutions.
Over time, teams with between-school collaborations consistently have greater expertise diversity than those formed within the same institution (Fig.~\ref{cit_diversity}{a}), and, similarly, international teams exhibit greater expertise diversity than domestic teams (Fig.~\ref{cit_diversity}{b}).
	
We find that for all three dimensions of external diversity regarding institutions, international collaborations, and gender, expertise diverse teams produce more disruptive work than teams with narrow expertise among coauthors (Fig.~\ref{cit_diversity}{d-f}).
This effect is stronger for small teams.
In particular, the work of teams that lack other dimensions of diversity tends to be more disruptive.
Papers written by teams from the same institute have higher disruption scores than between-school teams.
Similarly, domestic teams produce more disruptive science than those with international collaborations, and gender-homogeneous teams conduct more disruptive research than gender-diverse teams.
For teams that have other dimensions of external diversity and may therefore produce less disruptive work, increasing expertise diversity among team members appears to be particularly beneficial to enhancing the originality of research output.
	
Past studies suggest that between-school collaborations may facilitate knowledge production and encapsulate more external specializations, which promote high-impact research\cite{jones2008multi}.
We find that between-school teams have greater long-term paper-to-paper impact than teams formed within the same institution, for each specific team size and expertise distance percentile bin (Fig.~\ref{cit_diversity}{g}).
For large between-school teams with four or more authors, long-term impact increases as the teams' expertise diversity augments, but the effect is not obvious for small teams of two or three authors.
For teams of all sizes that are assembled in the same institution, long-term impact shows a remarkable boost as the team's expertise diversity increases, which becomes more pronounced as the team size grows.
In particular, for paper-to-paper citations, the long-term impact advantage of single-institutional teams in the highest distance percentile bin over those in the lowest distance percentile bin is $17.1\%$ for two-author teams, which rises to $49.0\%$ for teams of five or more authors.
For small teams of two to three authors, papers written by single-institutional teams have even higher impact on patents than those written by between-school teams (Fig.~\ref{cit_diversity}{j}).
Overall, while between-school teams garner more paper-to-paper citations than teams from the same institution, the impact upticks substantially faster for single-institutional teams as the teams' expertise diversity increases.
	
Previous research has also demonstrated that international collaborations promote impact, which we next investigate concerning the team's size and expertise diversity (Fig.~\ref{cit_diversity}{h})\cite{inzelt2009incremental, lancho2013citation, hsiehchen2015multinational, leydesdorff2019relative}.
We find that small- or medium-sized teams of two to four authors that have foreign collaborators do not exhibit a significant impact shift as expertise diversity increases, while large teams of five or more authors enjoy greater long-term impact if the team members' prior career histories are more diverse.
For papers written by single-country teams, however, bolstering the teams' expertise diversity is strongly associated with rising long-term impact, which becomes more pronounced as the teams' sizes grow.
The impact premium of single-country teams in the highest distance percentile bin over those in the lowest distance percentile bin is $16.0\%$ for two-author teams and $43.5\%$ for five or more author teams.
For small teams of two to three authors, domestic teams have even higher impact on patents than international teams (Fig.~\ref{cit_diversity}{k}).
Similar to the effect of between-school collaborations, an international team configuration obviously attracts more paper-to-paper citations, but domestic teams of all sizes enjoy greater boost of long-term impact as the teams' expertise diversity increases.
	
The above analyses regarding between-school and international collaborations suggest that the scientific and technological impact premium of teams with large expertise diversity is particularly pronounced for those formed within the constraints of geographical boundaries.
These findings are further validated in temporal time periods. 
Collaborating with researchers specializing in more remote areas appears to facilitate the production of high-impact work, especially for teams localized in their institutional or geographical spheres.
These findings thus suggest a plausible enhancement for teams that have limited alternatives of diversification and are restrained to mostly departmental colleagues or domestic collaborators.
For such teams, integrating the available expertise from a broad knowledge scope may predict higher-impact outputs.
	
Previous studies show that women are more inclined to step outside the disciplinary boundaries\cite{rhoten2007women}, more engaged in interdisciplinary collaborations\cite{van2011factors}, more likely to make scientific discoveries that lead to women-related health patents\cite{koning2021we}, and have a slightly higher propensity to collaborate with topically distant colleagues\cite{abramo2013gender,smith2021great}.
In the pre-2010 period, gender-diverse teams that include both men and women had greater expertise diversity than gender-homogeneous teams which are composed of men or women only, but gender-diverse teams have been paired with gender-homogeneous teams since the past decade (Fig.~\ref{cit_diversity}{c}).
We find less variation in the proportion of gender-diverse teams as a function of the team's expertise diversity after controlling for team size, compared to the previous institutional factors.
The propensity of having female collaborators increases slightly as the expertise distance of teams rises up to a moderate level, but declines in the highest distance percentile bin.
	
Gender-diverse teams perform equally well as gender-homogeneous teams in terms of long-term paper-to-paper citation impact when there are up to four authors, and significantly outperform gender-homogeneous teams when there are five or more authors (Fig.~\ref{cit_diversity}{i}).
Gender-diverse teams have higher long-term impact as their expertise diversity increases, and similar results hold for gender-homogeneous teams.
Promoting gender diversity in the team assembly process not only improves gender equality in science and provides more collaborative opportunities for women, but also enhances teams' performance in terms of high-impact research production\cite{yang2022gender}.
With respect to patent-to-paper citations, for both types of teams, increasing team expertise diversity is associated with greater impact on patenting, especially when the team has four or more authors.
However, gender-homogeneous teams still have moderately higher patenting impact than gender-diverse teams, especially when the team sizes are small (Fig.~\ref{cit_diversity}{l}).
Investigating the inherent causes for this gendered gap in patenting would facilitate the participation of women in industrial research, and make technology more diverse and vibrant in the knowledge-creating processes.

\section*{Discussion}
We propose a new method to quantify the expertise distance between two researchers based on the disciplinary distributions of their prior career histories. The method encapsulates several key aspects of expertise diversity, especially the intrinsic heterogeneity of interactions and relatedness among research fields. The expertise diversity of teams is strongly correlated with the teams' tendency to draw on broad combinations of past knowledge, disruption of produced work, long-term impact on research in other fields, and the spectrum of diversity among team members' affiliations and nationalities.
	
Although multidisciplinarity and team expertise diversity appear to reflect similar features of scientific diversity and both appear to be correlated with long-term impact, their correlations with disruption are considerably divergent.
Expertise diverse teams produce more disruptive work in science and technology, while high multidisciplinary inspiration exhibits no consistent association with disruption for papers and has negative correlations with disruption for patents.
Moreover, as recent decades have witnessed an explosive growth of newly created knowledge in science and technology, the rate of making major innovative progress and new directions of research is slowing in major disciplines\cite{chu2021slowed,park2023papers}.
The fact that high expertise diversity among team members tends to drive innovative research may provide a plausible solution to mitigate the slowing rates of disruption in science and technology.
Bringing together scholars with diverse expertise to incentivize original research would be especially viable for teams from less prestigious environments that have significantly less funding and resources compared to elite institutions.
	
Despite its correlation with multidisciplinarity, expertise diversity is essentially a distinct measure, and our work is not another study of interdisciplinarity.
By definition, multidisciplinarity measures how a work draws on the disciplinary scale of previous research, reflecting the diversity of knowledge used to develop a particular piece of work.
Quantifying and understanding the diversity of single entities is an inherently different computational task from measuring the disparity between multiple entities.
In contrast, expertise diversity reveals the breadth of knowledge scope among team members, which is essentially a measure to compare expertise disparity between two individual researchers.
In practice, an individual researcher alone can conduct highly multidisciplinary research, while expertise diverse teams can nevertheless decide to focus on narrow and specific research topics.
The difference between these measures is also manifested in their relation with disruption, as team expertise diversity exerts a strong boosting effect on disruption, a phenomenon not observed for multidisciplinarity.
	
While a team's expertise diversity has surprisingly little correlation with short-term or mid-term impact, teams with greater expertise diversity tend to garner substantially higher impact in the longer run. This trend becomes more prominent as the citation time window expands and the team size increases. Since most funding agencies assess research performance when funding ends (typically three to five years), these results raise the question of how best to evaluate the works of expertise diverse teams to capture their true potential to inspire future research.
	
Our distance measure sheds new light on the composition of prior expertise among team members, and offers predictive insights about key features of work the teams may produce, which may assist universities and funding agencies to identify and support highly diverse teams among the candidates\cite{chu2021slowed}.
Although teams with members from more diverse backgrounds, such as having between-school or international collaborators, usually outperform less diverse teams that are assembled within the same institution or country, our results suggest that pulling together a cohort of experts from a more diverse knowledge scope within the geographical constraints may be particularly powerful when external collaborators are unavailable.
Therefore, it raises the possibility that for teams that are restricted to only internal or domestic collaborators, preferentially recruiting more expertise diverse researchers might promote the long-term impact of the teams' work.
	
Note that the findings regarding expertise diversity of teams and their long-term impact are correlational and do not provide causal interpretations. 
It is also essential to exercise caution when interpreting disruption scores and consider them within the broader context of research impact and significance.
Combining methods in computational social science with tools from other areas, in particular, scientometrics, network science, and machine learning, could result in statistical and generative models that probe expertise diversity from broader perspectives. Efforts to explore underlying mechanisms that drive the formation of diverse teams, and factors that facilitate or hinder teams' long-term impact, are essential for researchers and policymakers to better understand the integration and production of knowledge in science and technology.

It is important to unpack hidden patterns to understand why expertise diverse teams garner more citations in the long run and identify the specific sources of these citations. 
In our current methodology of expertise diversity, we treat the distance between different disciplines as a continuous value.
It remains unclear how these disciplines are combined in the team formation process. 
Understanding the exact composition of field knowledge within teams and investigating which combinations of disciplines are most conducive to producing innovative research outcomes would be interesting topics for future research.

In line with previous studies, analyses regarding the expertise of individual researchers require decently long historical publication records to summarize their knowledge scope\cite{hill2021adaptability,zeng2021fresh}.
As such, in this study, we estimate expertise distance only for productive researchers that have at least 5 prior publications by the time of collaboration, and the team size refers to the number of such productive authors of a paper.
This constraint inevitably limits the availability of research papers and coauthor pairs to which empirical estimates of expertise distance are given, and thus only a fraction of teams and coauthors are eligible for the study of expertise diversity and its relation with disruption and impact.
Moreover, the method uses fields as its input parameter, thereby potentially influenced by the quality of field classifications.
	
The need to measure the similarity and diversity at the individual or system level is prevalent in many scientific, technological, and social systems\cite{stirling2007general,hidalgo2007product}. Our method provides a general framework of distance metrics that account for the relatedness of sub-categories within the system.
As such, this approach may have applicability in other settings, including but not limited to ecology diversity, innovation, research policy, and portfolio management, suggesting the potential for a more systematic exploration of similarity and diversity across broad domains.

\bibliography{scibib}

\end{document}